\documentclass[twocolumn,nofootinbib,floatfix]{revtex4-1}
\usepackage{graphicx}

\usepackage{amsmath,amssymb,fontenc, times, mathptmx}
\usepackage[dvips]{color}
\usepackage{latexsym,graphicx,epsfig}
\newcommand\blue{\color{blue}}

\def\lb{\linebreak[4]}
\newcommand{\be}{\begin{equation}}
\newcommand{\ee}{\end{equation}}
\newcommand{\bea}{\begin{eqnarray}}
\newcommand{\eea}{\end{eqnarray}}
\newcommand{\bes}{\begin{subequations}}
\newcommand{\ees}{\end{subequations}}
\newcommand{\bear}{\begin{equation}\begin{array}}
\newcommand{\eear}[1]{\end{array}\label{#1}\end{equation}}
\newcommand{\fr}[2]{\dfrac{{ #1}}{{ #2}}}

\newcommand{\fn}[1]{\footnote{{#1}}}

\renewcommand{\le}{\leqslant}

\def\vep{{\varepsilon}}
\newcommand{\epe}{\mbox{$e^+e^-\,$}}

\def\cl{\centerline}

{\end{list}}
\newcounter{enumct}
\newenvironment{Enumerate}{\begin{list}{\arabic{enumct}.}%
{\usecounter{enumct}\setlength{\topsep}{0.2mm}%
\setlength{\partopsep}{0.2mm}\setlength{\itemsep}{0.2mm}%
\setlength{\parsep}{0.2mm}\setlength{\leftmargin}{4mm}}}
{\end{list}}

\newcommand{\missET}{\slash{\hspace{-2.4mm}E}_T}

\begin{document}
\title{Singularities in  the single lepton energy spectrum for precise measuring  mass and spin of\\  Dark Matter particles  at the $e^+e^-$ Linear Collider}

\author{I. F. Ginzburg}
\affiliation{ Sobolev Institute of Mathematics and Novosibirsk State University,\\
{\it Novosibirsk, Russia}}


\begin{abstract}
{We consider models in which stability of  Dark Matter particles $D$  is
ensured by the conservation  of the new quantum number,  called
D-parity here. Our models  contain  also charged
$D$-odd particle $D^\pm$.

Here I propose the  method for precision measuring masses
and spin of $D$-particles via the study of energy distribution of
single lepton ($e$ or $\mu$) in the process $\epe\to
D^+D^-\to DDW^+W^-$ with the observable  states {\it dijet
+ lepton } ($\mu$ or $e$) + {\it nothing}. To  determine precisely masses of $D$ and $D^\pm$, it is sufficient to measure the singular points in the lepton
energy distributions (upper edge and kinks or peak). After this, even a rough
measuring of corresponding cross section allows to determine the spin of $D$ particles.

This approach is free from the difficulties of a well-known methods of measuring the masses via the edges of the energy distribution of  dijets, representing $W$, which obliged by inaccuracies in measuring the energies of individual jets.}
\end{abstract}
\pacs \, 95.30.Cq, 95.35.+d, 14.80.Ec, 14.80.Nb
\maketitle

\section{Introduction}\label{secintro}

We consider a wide class of models, in which Dark Matter (DM) consists of particles  $D$ similar to those in SM, with the following properties
(the examples are: MSSM where $D$
is  the lightest neutralino with spin $1/2$ \cite{MSSMdark},
and inert doublet model IDM \cite{inert} where $D$ is  the
Higgs-like neutral).
\begin{Enumerate}
\item DM particle $D$ with mass $M_D$ has  new conserved discrete quantum number. I call it  D-parity.  All known particles are
$D$-even, while the DM particle is $D$-odd (for MSSM $D$-parity means $R$-parity).
\item
In addition to the neutral DM particle  $D$,  another $D$-odd particles
exist,  a charged $D^\pm$ and (sometimes) a neutral $D^A$, with
the same spin $s_D=0 \mbox{ or } 1/2$ as $D$ and with masses
$M_\pm,\, M_A>M_D$. (In MSSM $D^\pm$ is  the lightest
chargino, $D^A$ is the second neutralino, in the IDM
$D^\pm$ is similar to the charged Higgs of 2HDM, $D^A$ is
similar to the CP odd scalar $A$ of 2HDM.) The D-parity
conservation ensures stability of the lightest $D$-odd particle $D$.

\item $D$-particles interact with the SM particles only via the Higgs boson $DDh$, $D^+D^-h$ and via the covariant derivative in the kinetic term of the Lagrangian -- gauge interactions with the standard electroweak gauge couplings { $g$, $g'$ and $e$ (for coupling to $Z$ -- with possible reducing mixing factor)}:
\end{Enumerate}
\be
 D^+D^-\gamma,\;\;  D^+D^-Z, \;\;  D^+DW^-, \;\; D^+D^AW^-,\;\; D^ADZ.
\ee

A possible value of mass $M_D$ is limited by  stability of
D-particles during the age of the Universe
\cite{Dolle:2009fn,PDG}. We will have in mind interval \be
4~\mbox{GeV}\lesssim M_D\lesssim 80~\mbox{GeV}\,. \ee

The non-observation of  processes $e^+e^-\to D^+D^-$ and
$e^+e^-\to  DD^A$ at   LEP gives   $M_+>
90$~GeV and limitation for $M_A$, dependent on $M_D$
\cite{Lundstrom:2008ai}. We assume below that mass difference
$M_+-M_D$ is not small, e.g. $>10$~GeV.

Experiments at  the Linear $e^+e^-$
Collider (LC), e.g. ILC/~CLIC, at $\sqrt{s}=2E>200$~GeV allow to
detect carefully the DM particle candidate and to  measure accurately  its
mass and  spin. In these tasks LC have many advantages as compared
with LHC.

{\bf Discovery}. The neutral and stable $D$ can  be produced and
detected  via  process with production $D^\pm$ or $D^A$ and
subsequent decay  $D^\pm\to DW^\pm$, $D^A\to DZ$ (with  either on shell or off shell gauge bosons $W$ and $Z$) , etc. To discover
DM particle, one  needs to specify such processes with clear
signature. As it is known (see e.g. \cite{TESLA}), the LC provides
excellent signature for such processes, see sect.\fn{In sect's
\ref{secsign+}--\ref{secmudistr} we consider the case when either
$D^A$ is absent or $M_A>M_+$, the case $M_A<M_+$ is considered in
sect.~\ref{seclDA}. } \ref{secsign+}, \ref{seclDA}, \ref{exppred}
-- note word {\it nothing} in \eqref{sign+W}, \eqref{signWZ1}.
Such signature is absent at LHC. Moreover, the cross section of
process $\epe\to D^+D^-$ is a large fraction of  the total cross
section of \epe annihilation.  At  LHC the cross section of
$D^+D^-$ production constitutes a small fraction of  the total
hadron cross section with large background +... Even
the separation of $q\bar{q}\to D^+D^-$ process at LHC is a
difficult task.

{\bf Masses}. The next problem is to determine   two  masses --
the "parental" (for example, $M_+$) and  the
"dark" $M_D$. For  this aim, it is
necessary to find in the kinematical characteristics of observed
particles at least 2 well separated points, measurable with good
precision, to have two equations for determination of $M_+$ and
$M_D$. Well known approach \cite{WILC} is to measure edges in the
energy distributions of dijets, representing $W$ from decay
$D^\pm\to DW^\pm$, sect.~\ref{secWdistr}. (For LHC similar
approach corresponds  to the study of edges in the
distribution of $M_T$ for dijets \cite{kinemLHC}). However,  the
individual jet energies and, correspondingly, effective mass of
the individual dijet cannot be measured  with high
precision. One can hope only to measure with satisfactory
precision the upper bound of energy distribution of $W$ in dijet
mode $E^{L,+}_{W}$ \eqref{EPW}, \eqref{ELWoff}, the lower
bound is smeared by uncertainty in the measuring of energy of an
individual jet. Therefore, such method cannot pretend for high
accuracy in the measuring of masses.

The lepton energy is measurable  with higher accuracy. However, in
the lepton mode of $W$ decay uncertainties, introduced momenta of
two invisible particles $D$ and $\nu$, make distribution of
leptons  more model dependent than that for $W$. Nevertheless, we
show in sect.~\ref{secmudistr} that  the energy distribution of
leptons has singular points which positions are kinematically determined, and -- therefore -- model independent.
Measuring  positions of these singularities
will allow to  determine masses $M_D$ and
$M_+$ with good precision.

Such simple opportunity is absent at LHC. Instead, at LHC one can try
to measure  the distribution of a single lepton in transverse momentum. At best, it will allow to measure one quantity (for example $p_\bot^{max}$),
which cannot give enough information about two masses $M_D$ and $M_+$.

{\bf Spin}. The cross section of process $\epe\to D^+D^-$ depends on $M_+$ and $s_D$ only, with  strong dependence on $s_D$ and weak dependence on detail of model. Therefore, after measuring of $M_+$ even rough measuring of cross section allows to select value of spin $s_D$ in model independent way. This is not possible at LHC, where production mechanism is model dependent. Here spin is either input parameter of model, or special measurements of more complex processes and distributions are necessary.

\section{Main process $\pmb{e^+e^-\to D^+D^-}$}

The energies, $\gamma$-factors and velocities of
$D^\pm$  are
 \be
 E_\pm=E=\sqrt{s}/2,\quad \gamma_+=E/M_+, \quad \beta_+= \sqrt{1-M_+^2/E^2}. \label{cmkinpm}
 \ee

\begin{figure}[hbt]\centering
\includegraphics[height=0.12\textheight,width=0.4\textwidth]{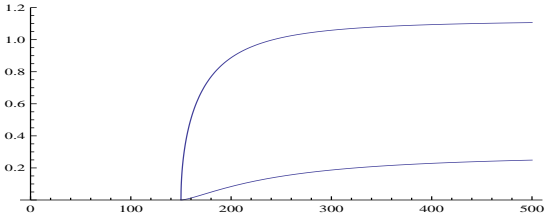}\hspace{5mm}
\caption{\it The $\sigma(e^+e^-\to D^+D^-)/\sigma_0$  dependence on $E$ at\\ $M_+=150$~GeV, upper curve -- $s_D=1/2$, lower curve -- $s_D=0$.}
\label{xsec+}
\end{figure}

Neglecting   terms
$\propto(1/4-\sin^2\theta_W)$, the cross section of process is a sum of model independent QED term (photon exchange) and $Z$ exchange term ( upper line -- for $s_D=1/2$, lower line -- for $s_D=0$):
\bear{c}
\!\!\sigma(e^+e^-\!\!\to\!\! D^+D^-)\!=\! \sigma_0\left\{\!\!\begin{array}{l}
\!\!\beta_+\!\left[\fr{3\!-\!\beta_+^2}{2}
+r_Z\beta_+^2\right]\!,\\[3mm]
\!\!\beta_+^3\!\left[\fr{1}{4}+r_Z\cos^2(2\theta_W)\! \right]\!
\!.\end{array}\right.\\[2mm]
r_Z=\fr{\mu_M}{\left(2\sin(2\theta_W)\right)^4(1-M_Z^2/s)^2}=\fr{0.124\mu_M}{(1-M_Z^2/s)^2}\,.
\eear{crsec}
Here $\mu_M\le 1$ is model dependent mixing factor, and
\be
\sigma_0\equiv \sigma(e^+e^-\to \gamma\to\mu^+\mu^-) =4\pi\alpha^2/3s\,. \label{sig0}
\ee
In Fig.~\ref{xsec+} and Table~\ref{tabA} we present these cross sections at $\mu_M=1$.

\begin{table}[hbt]
\caption{\it The cross section  $\sigma(\epe\to D^+D^-)$ for different spins}
\label{tabA}
{\begin{tabular}{|c||c|c|c|c|}\hline
$E$, GeV&100&250&250&250\\\hline
$M_+$, GeV & 80&80&150&200\\\hline
${s_D=0}:$ $\sigma/\sigma_0$&
0.066&0.245&0.162&0.062\\\hline
${s_D=1/2}:$ $\sigma/\sigma_0$&
0.84&1.107&1.02&0.82\\\hline
\end{tabular}}
\end{table}

Total cross section of the $e^+e^-$ annihilation at ILC for\lb
$\sqrt{s}>200$~GeV is  $\sim 10\;\sigma_0$. The cross section
\eqref{crsec} is $\sim \sigma_0$. Therefore, the  the number
of events of considered process is a significant fraction of all
the events  for $e^+e^-$ annihilation.

\section{$\pmb{M_A>M_+}$, signature}\label{secsign+}

 After  the
production, particles $D^\pm$ decay fast to $DW^\pm$
\be
\epe\to D^+D^-\to DDW^+W^-\,\label{DDWWmain}
\ee
with either
on shell (real) or off shell $W^\pm$, the latter is  $q\bar{q}$
pair (dijet) or $\ell\nu$, having the same quantum numbers as $W$
but effective mass $M^*<M_W$.
In both these cases the probability of this decay equals 1. The
observable states are decay products of $W$  with
large missing transverse energy \ $\missET$ carried away by the
neutral and stable $D$-particle + {\it nothing}, the
missing mass of particles escaping observation $M(\missET)$ is
large. Therefore, the  signatures of the process  in the modes, suitable for observation,  is
  \be
\boxed{\mbox{\begin{minipage}{0.37\textwidth} \cl{A) 2 dijets or} \cl{B) 1 dijet plus $ e$ or $\mu$}
\cl{with large   \  $\missET$  and large
$M(\missET)$  + {\large\it nothing},} total energy of each dijet or lepton less \
than  $E$.  \end{minipage} }}
\label{sign+W}
\ee

At $M^*>5$~GeV, the branching ratios (BR) for different
channels of $W$ decay are practically the same for on shell states
\cite{PDG} and off shell states.  In particular, the fraction
of events with 2 dijets from hadronic decays of both $W$'s \ \  is
$0.676^2\approx 0.45$. The fraction of events with 1 dijet from
$q\bar{q}$ decay of $W^\mp$  plus $\ell=\mu,\, e$ from lepton
decay of $W^\pm$  is $2\cdot 0.676\cdot2\cdot (1+0.17)\cdot
0.108\approx 0.33$ (here 0.17 is a fraction of $\mu$ or $e$ from
the decay of $\tau$).

At  $M^*<5$~GeV the BR's for
$e\nu$ and $\mu\nu $ modes increase while dijet degenerates into  set of few particles.

\section{$\pmb W$   energy distribution, $\pmb{M_A>M_+}$}\label{secWdistr}

Let us denote by $M^*$ the effective mass of  $q\bar{q}$ or $\ell\nu$ pair. At $M_+-M_D>M_W$ we have $M^*=M_W$ (on shell $W$), at\lb $M_+-M_D>M_W$ possible values of $M^*$ are within interval $(0,\, M_+-M_D)$
(off shell $W$). At each value of $M^*$ in the rest frame of $D^\pm$ we have 2-particle decay
 \bear{c}
\!E_{W*}^r\!=\!\fr{M_+^2 +M^{*2}- M_{D}^2}{2M_+},\quad p^r_{W*}\!=\!\fr{\Delta(M_+^2,M^{*2},M_{D}^2)}{2M_+},\\[2mm] \Delta(s_1,s_2,s_3)^2=s_1^2+s_2^2+s_3^2-2s_1s_2-2s_1s_3-2s_2s_3.
\eear{rkinW}

Denoting  by $\theta$ the $W^+$ escape angle in $D^+$ rest frame
with respect to the direction of $D^+$ motion in the Lab system
and using $c\equiv\cos\theta$, we find the energy of \ $W^+$ \ in
the Lab system  as $E_W^L=\gamma_+(E_{W*}^r+c\beta_+ p_{W*}^r) $.
Therefore, at given $M^*$ the energy  $E^L_W$  of $\ell\nu$ pair
or dijet from $W$ decay  lies within the
interval $\gamma_+(E_{W*}^r\pm \beta_+ p_{W*}^r)$.

At  ${M_+-M_D>M_W}$ we deals with on shell $W$, and  this equation
describes kinematical edges of $W$ energy:
 \be
E^{L,\pm}_{W,on}\!=\!\gamma_+(E_W^r(M_W)\!\pm\!\beta_+
p^r_W(M_W)).\label{EPW}
 \ee

At  ${M_+-M_D<M_W}$ similar edges are different  for each value of
$M^*$. In particular, at the  highest value $M^*=M_+-M_D$ we have
$p_W^r=0$, and interval, similar to \eqref{EPW} reduces to a point, where entire $W$ energy distribution has maximum (peak)
\be
E^L_{W,p}\equiv
E^{L, \pm}_W|_{(M^*=M_+-M_D)}=E\left(1-\fr{M_D}{M_+}\right).\label{peakW}
 \ee
Absolute upper and lower bounds on the energy distribution of the
muons are achieved at $M^* = 0$,  they are
 \be
E^{L,\pm}_{W,off}=
E\,\fr{1\pm \beta_+}{2}\left(1-\fr{M_D^2}{M_+^2}\right)
\;.\label{ELWoff}
\ee

\section{Single lepton  energy distribution in $\pmb{\epe\to D^+D^-\to DD W^+W^-\to DD q\bar{q}\,\ell\nu}$}\label{secmudistr}

The lepton energy $\vep$ is measurable with high accuracy. Therefore it is useful to study  the energy distribution\fn{Here we include arguments, marked masses, responsible for the form of distribution. } $d\sigma_0^\mu(\vep|M_+,\,M_D)/d\vep $ for the events with signature (\ref{sign+W}B) more attentively.  We find that this distribution  has singular points which positions are model independent. We  consider,
for\ \  definiteness, $\ell=\mu^-$, neglect the muon mass and limit ourself in this section to the case $M_A>M_+$.

a) If ${M_+-M_D>M_W}$, the muon energy and momentum in the
rest frame of $W$ are $M_W/2$. In the Lab system for $W$ with some energy $E_W^L$ the  $\gamma$-factor and the
velocity of $W$ are\lb $\gamma_{WL}=E_W^L/M_W$ and
$\beta_{WL}\equiv \sqrt{1-\gamma_{WL}^{-2}}$. Just as above,
denoting by $\theta_1$ the escape angle of $\mu$ relative to the
direction of the $W$ in the Lab system and $c_1=\cos\theta_1$, we
find that in the Lab system the muon energy $\vep=\gamma_{WL}\left(1+c_1\beta_{WL}\right)(M_W/2)$. Therefore
$$
\vep^+(E^L_W)\geqslant \vep \geqslant \vep^-(E^L_W)\equiv M_W^2/\left(4\vep^+(E^L_W)\right),
$$
where $ \vep^+(E^L_W)= E_W^L\fr{1+\beta_{WL}}{2}=\fr{E_W^L+\sqrt{(E_W^L)^2-M_W^2}}{2}
$.

\begin{figure}
\includegraphics[height=0.18\textheight,width=0.55\textwidth]{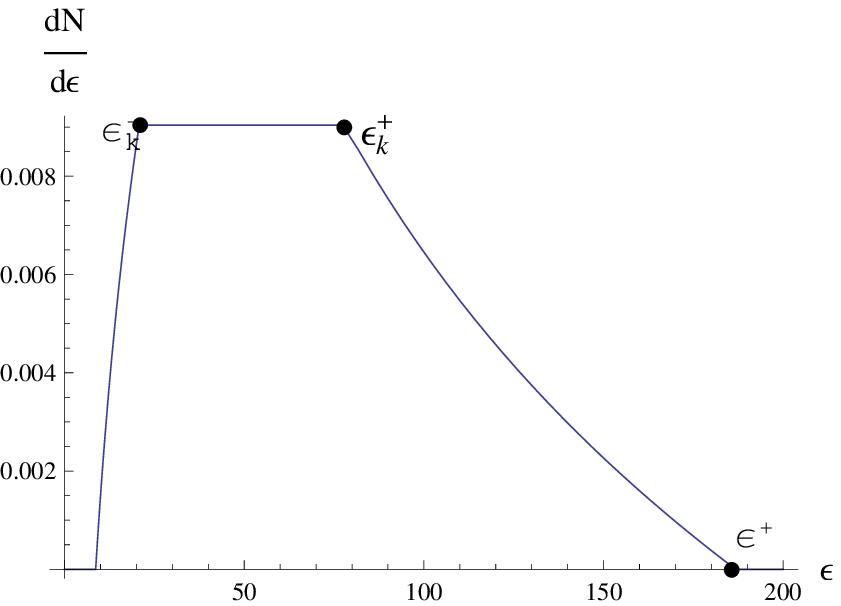}
\includegraphics[height=0.2\textheight,width=0.48\textwidth]{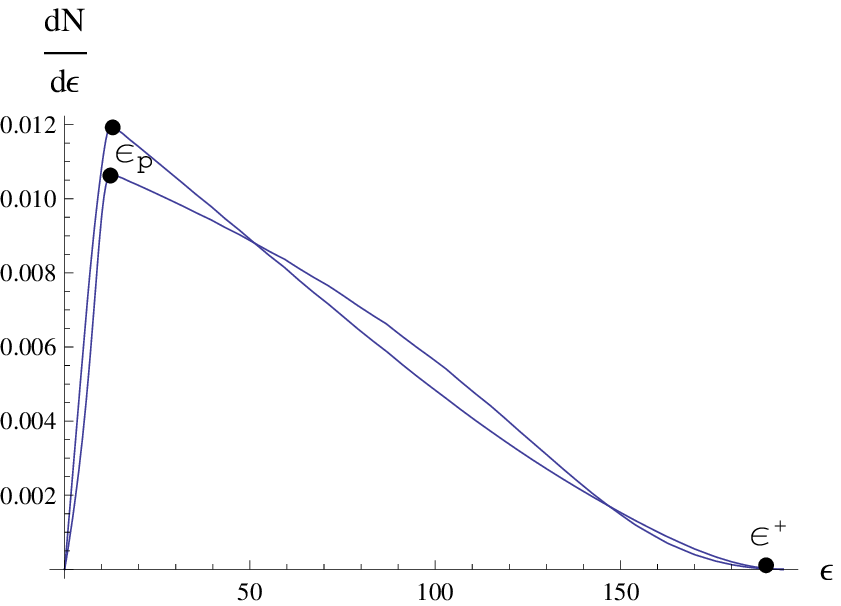}
\caption{The normalized  distributions $dN/d\epsilon\equiv
(1/\sigma)d\sigma/d\epsilon$ for $M_+=150$~GeV  (on shell $W$) -- upper plot,
for $M_+=120$~GeV (off shell $W$) -- lower plot,     $E=250$~GeV.
At lower plot upper peak -- for $s_D=0$, lower peak -- for
$s_D=1/2$.} \label{singlmufig}
\end{figure}

The interval,  corresponding to energy
$E_{1W}^L<E_W^L$, is located entirely within the interval,
correspondent to  energy $E_W^L$. Therefore, all muon energies lie
within the interval determined by the highest value of $W$ energy:
 \bear{c}
 \vep^+ \geqslant \vep\geqslant \vep^-\equiv \fr{M_W^2}{4\vep^+},\;\;where\\[2mm]  \vep^+\equiv
\vep^+(E^{L, +}_{W,on})=
 \fr{
E^{L, +}_{W,on}+\sqrt{(E^{L, +}_{W,on})^2-M_W^2}}{2}\,.
\eear{Emularge} Contributions of  $W$ with intermediate energies
are summarized in the entire distribution of muons in the energy,
and it increases monotonically from the outer limits to kinks at
energies $\vep^\pm_k$, corresponding to the lowest
energy $E^{L,-}_{W,on}$  of the $W$ boson:
 \be
  \vep^\pm_k\equiv
\vep^+(E^{L, -}_{W,on})=
 \fr{
E^{L, -}_{W,on}\pm \sqrt{(E^{L, -}_{W,on})^2-M_W^2}}{2}\,.\label{Emuin}
 \ee
Between these kinks $d^2N/d\vep^2 \approx 0$. The  energy
distribution of muons for the case of matrix element, independent
on $\theta_1$, is shown in  Fig.~\ref{singlmufig} -- up. Calculations for separate models (where angular dependence exists)  demonstrate variation in details of shape of these curves but the position of kinks is fixed \cite{GKr}.

b) If ${M_+-M_D<M_W}$, the $D^\pm$ decays to $(D+W^*)$ where $W^*$ is off
shell $W$ with effective mass $M^*\leqslant
M_+-M_D$. The calculations, similar to above, for each $M^*$  shows \ that the muon energies  are within the interval, appearing at $M^*=0$:
 \be
\left\{\vep^-=0;\;\vep^+=
E\,\fr{1+ \beta_+}{2}\left(1-\fr{M_D^2}{M_+^2}\right)\equiv E^{L,+}_{W,off}
\,\right\}.
 \label{emustarmax}
 \ee
Similarly to the preceding discussion, the increase of $M^*$
 shifts the interval boundaries
inside. Therefore, the muon energy distribution increases
monotonically from outer bounds up to the maximum (peak) at
$M^*=M_+-M_D$ (cf. \eqref{peakW}):
\be
\vep_p=E\fr{1+\beta_+}{2}\left(1-\fr{M_D}{M_+}\right).
 \label{emustarmin}
 \ee

To get an idea about the shape of the peak, one should use the
distribution of $W^*$'s (dijets or $\ell\nu$ pairs)  over the effective masses $M^*$. It is given by the spin
dependent factor $R_{s_D}p^*dM^{*2}$:
 \bear{c}
R_0=\fr{p^{*2}}{(M_W^2-M^{*2})^2}\,,\\
\!\!\!\!R_{1/2}\!=\!\fr{(M_+^2\!+\!M_D^2\!-\!M^{*2})(2M_W^2\!+\!M_+^2\!+\!M_D^2)\!-\!4M_+^2M_D^2}{(M_W^2-M^{*2})^2M_W^2}.
\eear{offshelmass}

The density of muon states in energy $dN/d\vep$ is calculated by
convolution of kinematically defined distribution with
distribution \eqref{offshelmass}.  Neglecting the dependence of
the matrix element of the angle, we obtain result in form of
Fig.~\ref{singlmufig}-down. One can see that the discussed peak is
sharp enough for both values of spin $s_D=0$ and $1/2$.

Characteristic values for singular point (kink and peak) energies
in these distributions (together with similar points for energy distributions of $W$ (dijets)) are given in the table~\ref{tab+} .
\begin{table}[htb]
\caption{\it The singular point energies of lepton and $q\bar{q}$ dijet
in\\ $\epe\to D^+D^-\to DDq\bar{q}\ell\nu$  (in GeV) at $M_D=50$~GeV. }\label{tab+}\vspace{2mm}
\begin{tabular}{|c|c||c|c|c| c|c|}\hline
$E$ &$M_+$&$\vep^+$&$\vep^+_k $&$\vep_p$&$E^L_{Wp} $&$E^{L,+}_W$\\ \hline
250&150&186.3&77.8&-&-&195.4\\ \hline
250&200&184.9&46.3&-&-&193.6\\ \hline
250&80&148.3&-&91.3&93.75&148.3\\\hline
100&80&78&-&30&37.5&78\\\hline
\end{tabular}
\end{table}


{\bf The cascade $\pmb{D^+\to DW^+\to D\tau^+\nu\to
D\mu^+\nu\nu\nu}$} modifies  spectra under
discussion. The energy distribution of $\tau$, produced in the
decay $W\to \tau\nu$, is the same as that for $\mu$ or $e$,
discussed above (with accuracy $M_\tau/M_W$ or $M_\tau/M^*$).
After its production, $\tau$ decays to $\mu\nu\nu$ in 17 \% cases
(the same for decay to $e\nu\nu$). These muons are added to the
above discussed.

In the $\tau$ rest frame the energy of muon
$E_\mu^\tau=y\,M_\tau/2$ with $y\leqslant 1$. The energy spectrum
of muons  is $dN/dy=2(3-2y)y^2$ (see textbooks). This
spectrum and distributions, obtained above, are converted into the
energy distribution of these muons in the Lab system. Two features
of this contribution are clear  on the qualitative level

A) This contribution is shifted strong to the soft part of energy spectrum.

B) This contribution has no singular points with jump of derivative in $\vep$.

{\it The resulted muon energy distribution} is similar to that
without $\tau$ contribution, Fig.~\ref{singlmufig}. This
contribution does not change the upper end point of the energy
distribution of the muons $\vep^+$ \eqref{Emularge},
\eqref{emustarmax}.  Numerical examples \cite{GKr} show that the
discussed correction shifts positions of kinks or
peak in the muon energy distributions by less than  1 GeV,
i.e. negligibly.

\section{Case $\pmb{M_+>M_A}$}\label{seclDA}

For the main process $\epe\to D^+D^-$ at $M_+>M_A$ {\bf one more
decay become also possible, $\pmb{D^\pm\to D^A W^\pm\to DZW^\pm}
$}. Total probability of decays $D^+$ to $D^AW^+$ and  $DW^+$
equals 1. The decay $D^\pm \to D^AW^\pm$ is described by the same
equation as $D^\pm \to DW^\pm$, but with another kinematical
factors since $M_A> M_D$. The probability of this new decay is
lower than that without $D^A$ due to smaller final phase space
volume, i.e. $B= BR(D^+\to D^AW^+)<0.5$.

In the same manner as above, particle $D^A$ decays fast to $DZ$
and we deals with cascades\\ \cl{$e^+e^-\to D^+D^-\to
DW^+D^AW^-\to DD W^+W^-Z$, etc.} Now signature of  processes
$e^+e^-\to D^+D^-$  in the modes, suitable for observation,
contains both \eqref{sign+W} and \be
\boxed{\mbox{\begin{minipage}{0.41\textwidth}  3 or 4 dijets, or
less dijets plus 1 to 5 leptons with large    $\missET$  and large
$M(\missET)$  + {\large\it nothing}.   \end{minipage}
}}\label{signWZ1}
 \ee

Note that 20\% of final states of $Z$ decay are invisible
($\nu\bar{\nu}$ final states). We denote these states
as $Z_n$.

Let us consider in more detail final states with signature
(\ref{sign+W}B) (observed state:  1  dijet $+\mu^-$ +{\it
nothing}). This state can be obtained from two group of channels with different
mechanism of cascades $D^-\to D\mu^-+\ldots$ \ \
and all possible channels for decay $D^+$:\\
1) Channels where $D^-$ decays to $DW^-\to D\mu^-\nu$.
The energy distribution of $\mu^-$ in these channels reproduces that,
obtained for the case $M_A>M_+$ (Sect.~\ref{secmudistr}), that
is\lb  $(1-B) d\sigma_0^\mu(\vep|M_+,\,M_D)/d\vep $.
Here  $d\sigma_0^\mu(\vep|M_+,\,M_D)/d\vep $ is energy distribution
obtained for the case $M_A>M_+$, we have written explicitly the
arguments indicating mass  of the initial and final $D$-particles.\\
2) Channels where $D^-$ decays to $D^AW^-\to DZ_n\mu^-\nu$. Since
couplings $D^-DW^-$ and $D^-D^AW^-$ differ by  phase
factor only, the energy distribution of $\mu^-$ in these
channels  is described by the same dependence $d\sigma_0$ but with
the change $M_D\to M_A$, the corresponding contribution to the
entire energy distribution is $0.2B
d\sigma_0^\mu(\vep|M_+,\,M_A)/d\vep $. For brevity we will write
$d\sigma_0^\mu(\vep|M_+,\,M_D)\to d\sigma^\mu_W$ and
$d\sigma_0^\mu(\vep|M_+,\,M_A)\to d\sigma^\mu_{WZn}$. The
resulting energy distribution is
\be
d\sigma_{tot}^\mu/d\vep=(1-B)d\sigma_W^\mu/d\vep
+0.2Bd\sigma_{WZn}^\mu/d\vep\,. \ee

The shape of distribution $d\sigma_{WZn}^\mu/d\vep$ is similar to
that for $d\sigma_W^\mu/d\vep$ (Sect.~\ref{secmudistr}) but with
another positions of kinks and (or) peak. Since $M_A>M_D$, these new kinks
and (or) peak are situated below similar positions for
$d\sigma_W^\mu/d\vep$. Since this contribution is much smaller
than the main contribution $d\sigma_W^\mu/d\vep$ (with
overall ratio $0.2B/(1-B)$ at $B<0.5$), it results in only weak
change of entire energy distribution as compare  with
distributions   in Sect.~\ref{secmudistr}. The
opportunity to extract from the data new singularities, related to
$d\sigma_{WZn}^\mu/d\vep$, demands separate study.

\section {Discovery,  measuring of masses and spin}\label{exppred}

{\bf Discovery}. The observation of  events with signature
\eqref{sign+W}, \eqref{signWZ1} will be a clear {\it
signal} of candidates for DM particles. The process with
signature \eqref{signWZ1} can take place only simultaneously with
processes $\epe\to DDZ$ with signature \eqref{DDAZ}.

{\bf Masses $\pmb{M_+}$ and $\pmb{M_D}$} can be determined from singular points of the energy distribution
of the leptons in the final state  $q{\bar q}\ell$  + {\it
nothing} by summing contributions  from $e$ and $\mu$. With
anticipated annual luminosity integral $\cal L$ for  the ILC
project \cite{WILC} ${\cal L}\sigma_0\sim 10^5$
the 1-year number of events of this type will be $\sim (1\div
3)\cdot 10^4$, depending on  masses and spin $s_D$.

{\bf M1)} If $D^A$ particle is absent or  at $M_+<M_A$, the
results of Sect.~\ref{secmudistr} describe the energy
distributions completely. The shape of energy distribution of
leptons (with one peak or two kinks) allows to determine
what case is realized,\lb $M_+-M_D>M_W$ or $M_+-M_D<M_W$. At
$M_+-M_D>M_W$ the position of upper edge of the muon energy
$\vep^+$ \eqref{Emularge} and one kink, e.g. $\vep^+_k$
\eqref{Emuin} give us two equations necessary for
determination of $M_D$ and $M_+$.  At $M_+-M_D<M_W$
two  similar equations are given by the position of
upper end point of the muon energy $\vep^+$ \eqref{emustarmax} and
peak $\vep_p$ \eqref{emustarmin}.

The singular points of dijet energy distribution  can be also used for measuring on masses.

At $M_+-M_D>M_W$ the upper edges of dijet energy distribution $E^L_W$ and muon energy distribution $\vep^+$ contains  identical information, since $E^{L,+}_W\!\!=\!\!\vep^+\!\!+\!M_W^2/4\vep^+$ (cf.~\eqref{EPW}, \eqref{Emularge}). In this  case results of measuring $E^{L,+}_W$ and  $\vep^+$ supplement each other.

At $M_+-M_D< M_W$ we have $E^{L,+}_W=\vep^+$ at $M_+-M_D<M_W$ (cf.~\eqref{ELWoff}, \eqref{emustarmax}). In this case measuring of $E^{L,+}_W$ meet additional difficulties since this upper edge is given by values of $M^*$, close to 0, when dijet is degenerated into 2-3 pions. The position of peak in the dijet  energy distribution $E^L_{Wp}$ looks useful since $\vep_p/E^L_{Wp}=(1+\beta_+)/2$ (cf.~\eqref{peakW}, \eqref{emustarmin}). However position of this peak in the dijet distribution
is smeared by an uncertainty in the  measurement of the energy of individual jets.

{\bf M2)} For the case $M_+>M_A$  the entire energy
distribution of muons in the observed state
{\it $\mu$ +1 dijet + nothing}  was described  in
Sect.~\ref{seclDA}.  As it was mention there, taking into account
a new decay channel $D^-\to D^AW^-\to DZ_n\mu^-\nu $
changes  the position  of  the main singularities in
the muon energy spectrum only a little. Therefore the above
mentioned procedure for finding $M_+$ and $M_D$ can be used in
this case as well.

Note that in the case
$M_A\approx M_D$ distributions
$d\sigma_{WZn}^\mu/d\vep$ and $d\sigma_W^\mu/d\vep$ are close to
each other, and discussed procedure describes "degenerated" quantity $M_A\approx M_D$. In the opposite degenerate case
$M_+\approx M_A$ quantity $B\ll 1$, and influence of
intermediate $D_A$ state  on the result is
negligible.

{\bf M3)}  At $M_A+M_D<2E$
the process
\be
e^+e^-\to Z\to DD^A\to DDZ \label{DDAproc}
\ee
 becomes possible with clear signature
\be \boxed{\mbox{\begin{minipage}{0.41\textwidth} The  dilepton
(\epe \ or $\mu^+\mu^-$ \ pair)  or quark dijet with large \
$\missET$  and large $M(\missET)$ + {\large\it nothing}. The
effective mass of this dilepton or dijet is either $M_Z$ or lower
than $M_Z$. \end{minipage} }} \label{DDAZ}\ee The cross section of
this process is also $\sim \sigma_0$ but it is
smaller than that for production $D^+D^-$ \eqref{crsec} with
smaller BR for lepton mode. Moreover, the value of this cross
section is highly model dependent. With annual luminosity
\eqref{sig0}, the 1-year number of events of this type will
be $\lesssim (3\div 15)\cdot 10^2$ (depending on   masses, spin
$s_D$ and details of the model) \cite{Gin10}.

The calculations similar to those for $W$ energy distribution for
process \eqref{DDWWmain} allow to obtain kinematical edges of
the energy distribution of dilepton for each value
of its effective masses like \eqref{EPW}-\eqref{ELWoff}. Measuring
these edges gives two equations for finding $M_A$ and
$M_D$. (If $M_A-M_D<M_Z$,
this procedure must be performed separately for each value of the effective mass of dilepton.)  \cite{Gin10}, \cite{WILC}, \cite{kinemLHC}.

{\bf Spin of $\pmb D$-particles $\pmb{s_D}$}.  The  cross section
of the process $e^+e^-\to D^+D^-$ is obtained by summation over
all processes with signature \eqref{sign+W}, \eqref{signWZ1}
taking into account the known BR's for $W$ decay.

When masses $M_+$ become known, the cross section of the process
$e^+e^-\to D^+D^-$ is calculated easily for each value of spin
 \eqref{crsec}. The main part of the $\sigma(e^+e^-\to D^+D^-)$ is given by model
independent QED contribution of photon exchange, whereas the model
dependent contribution of $Z$ exchange at $\sqrt{s}>200$~GeV
contributes less than 30\%.  For identical masses
$\sigma(S_D=1/2)>4\sigma(s_D=0)$ (cf. table~\ref{xsec+} and
Fig.~\ref{xsec+} for examples).  This strong difference in the
cross sections for different $s_D$ allows to determine spin of $D$
particle even at low accuracy in the measuring
of cross section.

The similar procedure for the process $\epe\to DD^A$ cannot
be developed  in the model independent way due to
the strong model dependence of cross section.

\section{Background}\label{bg}

$\pmb {BW1}$. The process $\pmb{\epe\to W^+W^-}$ gives the same
final state as our process \eqref{sign+W}. However,   many of its
features are not permitted in signature
\eqref{sign+W}.\\
(a) Energy of  each dijet equals $E$.\\
(b) For the dijet+dijet observable state the observed
\, $\missET$ is low (in an ideal case\, $\missET=0$).\\
(c) For the dijet +lepton  state the missing mass $M(\missET)$ is
low (in  an ideal case $M(\missET)=0$.\\
These differences allow to exclude process $BW1$ from {\blue the}
analysis with  a good confidence by application of  suitable cuts.

$\pmb{BW2}$. $\pmb{\epe\to DD^A\to  DD^+W^-\to DDW^+W^-}$  {\bf
at} $\pmb{ M_A>M_+}$. If $\sigma(\epe\to DD^A)$ is not small at
given $\sqrt{s}$, this fact will be seen via observation of the process  $\epe\to DDZ$ \eqref{DDAZ}. The cross section
$\sigma(BW2)< \sigma(\epe\to DDZ)$, i.e.  it is much less than
$\sigma(\epe\to D^+D^-\to DDW^+W^-)$. Its  contribution may be
reduced additionally by application of cuts
taking into account the following points. \\
(a) In the process $BW2$ all recorded particles move in one
hemisphere in contrast with process \eqref{sign+W}, where they
move in two opposite hemispheres.\\ (b) In the process $BW2$ total
energies of lepton and jet are typically very different in
contrast to the process\linebreak[4] $\epe\to W^+W^-\to DDW^+W^-$
where these energies are close to each other.

$\pmb{BW3}$. {\bf In the   SM processes with observed state,
satisfying criterion \eqref{sign+W}}, large $\missET$ is carried
away by additional  neutrinos. The corresponding cross section is
at least one electroweak  coupling constant squared $g^2/4\pi$ or
$g^{\prime 2}/4\pi$ smaller than $\sigma_0$, with $g^2/4\pi\sim
g^{\prime 2}/4\pi\sim \alpha$. Therefore, the cross sections for
these  background processes are by about one or two orders
of magnitude smaller than the cross section of the process under discussion.\\

We discuss also briefly background processes for $\pmb{e^+e^-\to
DD^A\to DD \ell^+\ell^-}$. These processes are subdivided  into 3 groups.

{\bf BZ1.}  $\epe \to ZZ_n$.   At first sight, this  process can
mimic the process $\epe\to DDZ$. However, the lepton or
quark pairs in the process BZ1 have the same energy $E$ as the
colliding electrons. Therefore the criterion \eqref{DDAZ} excludes
such events from the analysis.

The cross section  $\sigma(\epe\to ZZ_n)\sim 0.2\cdot 3r_Z\sigma_0
\ln (s/M_Z^2)$. The variants of this process with off shell $Z$,
giving another effective mass of observed dijet or dilepton and,
respectively, another values of their energy, has cross
section which is smaller by factor $\sim \alpha$.

{\bf BZ2.} Processes with independent production of separate:\\
(BZ2.1) $\epe\to DDZ\to DD\tau^+\tau^-\to DD \,
\ell_1^+\ell_2^-+\nu's$, \\
(BZ2.2) $\epe\to DD^A\to DDW^+W^-\to
DD\ell_1\bar{\ell}_2\nu\bar{\nu}$,\\
(BZ2.3) $\epe\to D^+D^-\to DDWW\to
DD\ell_1\bar{\ell}_2\nu\bar{\nu}$,\\
(BZ2.4) $\epe\to W^+W^-\to \ell_1\bar{\ell}_2\nu\bar{\nu}$.\\
In these processes $\epe$, $\mu^+\mu^-$, $e^-\mu^+$ and
$e^+\mu^-$ pairs are produced with identical probability and
identical distributions. Hence,
 \bear{c}
\mbox{\it the subtraction from the $\epe$ and $\mu^+\mu^-$ data  }\\
\mbox{\it the measured distributions of \ $e^-\mu^+$ and \
$e^+\mu^-$} \eear{subtr} eliminates contribution of these
processes from the energy distributions under interest. This
procedure  does not implement substantial
inaccuracies since cross sections of these processes after
suitable cuts  will be small enough.

The cross sections of processes (BZ2.1), (BZ2.2) are  small in
comparison with that for $e^+e^-\to DD\mu^+\mu^-$. In the process
(BZ2.3) leptons are flying in the opposite hemisphere, in contrast
to the process under study $\epe\to  DDZ\to DD\mu^+\mu^-$, where
the leptons are flying in the same hemisphere The  cross section
of the process (BZ2.4) is basically large. The application
of cuts $E_{\ell\bar{\ell}}<E$, $M_{\ell\bar{\ell}}\le M_Z$
leaves  less than $(M_Z^2/s)^2\ln(s/M_Z^2)$ part of the cross section. The obtained quantity becomes smaller than that for the signal.

{\bf BZ3. In the  SM processes with observed state
\eqref{DDAZ}}, the large $\missET$ is carried away by
additional  neutrino(s). The magnitude of corresponding cross
sections are at least by one electroweak
coupling constant squared $g^2/4\pi$ or $g^{\prime 2}/4\pi$ less
than $\sigma_0$, with $g^2/4\pi\sim g^{\prime 2}/4\pi\sim \alpha$.
Therefore, the cross sections of these processes are at least one
order of magnitude smaller than the cross section
for the signal process.

{\bf Some limitation.}

In the real analysis, the energy spectra under
discussion will be smeared due to
initial state radiation and beamstrahlung.\\

This work was supported in part by grants RFBR 11-02-00242,
NSh-3802.2012.2, Program of Dept. of Phys. Sc. RAS and SB RAS
"Studies of Higgs boson and exotic particles at LHC" and  Polish
Ministry of Science and Higher Education Grant N N202 230337. I am
thankful A.E. Bondar, A.G. Grozin,  I.P. Ivanov, D.Yu.~Ivanov,
D.I.~Kazakov, J.~Kalinowski, K.A.~Kanishev, P.A.~Krachkov and
V.G.~Serbo for discussions.

\end{document}